\documentstyle[12pt]{article}
\begin{document}
\begin{center}
\vskip 0.001in
{\Large \bf Cross-over from BCS superconductivity to Bose condensation
and high-$T_c$ superconductors}
\vskip 0.4in
{\bf
Georgios VARELOGIANNIS}
and {\bf Luciano PIETRONERO}
\vskip 0.15in
{\em Dipartimento di Fisica,
     Universita di Roma "La Sapienza"\\
     Piazzale Aldo Moro 2, I-00185 Roma, Italy}
\vskip 0.15in
\begin{abstract}
\baselineskip 0.75cm
We consider the Eliashberg theory in the
coupling region where
some fundamental qualitative deviations from
the conventional BCS-like behaviour begin to appear.
These deviations are identified as the onset
of a cross-over from BCS
superconductivity to Bose condensation.
We point out that
the beginning of this cross-over
occurs when
the gap $\Delta_g$ becomes comparable
to the boson energies $\Omega_{ph}$.
This condition is equivalent to the condition of Ref. \cite{Strinati}
$k_F\xi\approx 2\pi$ and
traduces the physical
constraint that the distance the paired
electron covers during the absorbtion of the virtual boson, cannot be
larger than the coherence length.
The frontier region of couplings
is of the order of $\lambda\approx 3$,
and high-$T_c$ materials are concerned.
A clear qualitative indication of the occurence of a
cross-over regime should be a dip structure above the gap in the density
of states of excitations.
Comparing our results with tunneling and photoemission experiments
we conclude that high-$T_c$ materials (cuprates and fullerides) are indeed
at the beginning of a cross-over from BCS superconductivity to Bose
condensation, even though the fermionic nature still prevails.
Taking into account the analysis of Ref. \cite{Strinati}, we predict
a dip structure
in heavy fermion and organic superconductors.
Non-adiabatic effects beyond Migdal's theory are considered
and give insight on the robustness of Eliashberg theory in
describing qualitatively this cross-over regime, although for
the quantitative interpretation of the results the inclusion
of non-adiabatic corrections can be important.

\end{abstract}
\vskip 0.3cm
PACS numbers: 74.25.-q, 74.72.-h, 74.70.Wz \hfill
\end{center}
\newpage
\baselineskip 0.85cm

\vskip 0.6cm
\begin{center}
{\bf I. INTRODUCTION}
\end{center}
\vskip 0.4cm

Since the discovery of high-$T_c$ superconductivity a large effort has
been made to understand its origin but the question remains unsolved
and highly controversial.
Although
attempts have been
made
to fit the high-$T_c$ superconducting properties
within the conventional Eliashberg theory (ET) framework
\cite{Germans,Ruses,Canada,Zeyher,Japs}, many people believe that the
relevent mechanism
might be new based also on the idea that the critical
temperatures accessible by conventional theories should
exceed  $\approx 20K$ \cite{Elbio}.
There is not rigourous justification
for this idea which is due principally to an erroneous
conclusion based on partial numerical results by
McMillan \cite{McM}, corrected a long time ago by
Allen and Dynes \cite{AlDy}.

Actually,
there is not formal limitation on the critical temperatures one can
reach within ET provided the electron-boson coupling
strength or mass enhancement parameter $\lambda$ is
sufficiently strong. However it is not clear which are the maximal couplings
$\lambda$
for which ET is valid. Indeed as the coupling
strength grows the system should leave the BCS pairing regime
and switch to a Bose condensation regime \cite{Benoy,AR1,MRR}.
This last regime is considered by
several authors as relevant in relation to
high temperature superconductivity \cite{AR2,Kami,Mott}.

Little is known on the conditions under which this cross-over occurs.
Several authors have investigated the intermediate regime between BCS pairing
and Bose condensation for particular models
\cite{Noz,rand,Ciuchi,Zwerger,Avignon,Strinati}.
For a most general treatement of this very important intermediate regime
it is necessary to extend Eliashberg theory beyond Migdal's
theorem. Although significant progress have been
achieved in this last direction \cite{LP,LP2,D2,Gunna2}
the problem remains extremely complicate.

It is surprising that a study of ET at $\lambda\rightarrow\infty$
(where of coarse ET is not valid) leads
to a regime with local pairs \cite{MCinf},
however quantitatively
this regime is very different from the realistic Bose condensation regime
\cite{MRR}. This is an important remark that points to an unexpected
robustness of ET.
Here we investigate the question of the cross-over
from BCS superconductivity to Bose condensation from the ET
point of view \cite{Elias}. We will be concentrated
in the region of couplings where we first identify signs of
a cross-over regime.
We will see in particular
which are the couplings for which the need to go
beyond Migdal's theorem is manifest within ET.
To this fundamental question there is not
clear answer up to now. We will try also to locate the
high-$T_c$ materials with respect to this cross-over.

This paper is organized as follows.
In section II we study the structure of the gap function and point out
that the cross-over should occur when the gap $\Delta_g$ becomes
of the order of the phonon frequency $\Omega_D$. In section III
we give a simple physical interpretation of this condition
and show its equivalence to the condition derived in Ref. \cite{Strinati}
from a different perspective. In section IV we discuss the
location of high-$T_c$ SC with respect to this crossover.
In particular we analyze the properties of the density of states
and we conclude that they are indeed close to this crossover but
still preserve a fermionic character. In section V we discuss the
robustness of Eliashberg theory by including corrections
beyond Migdal's theorem. Finally in section VI we discuss the
conclusions of this work and outline the possible developpements.

\vskip 0.6cm
\begin{center}
{\bf II. CROSS-OVER AND THE GAP FUNCTION}
\end{center}
\vskip 0.4cm

When the characteristic energies of the superconducting state
(gap, $T_c$) are not completely negligible compared to the
boson energies (which are the characteristic energies for the
variations of the propagators in the coupled electron-boson
problem) for the correct description of the
BCS-type superconductivity it is necessary to consider the
retarded nature of the effective interaction and to use the
more elaboarte ET framework. Although initially this framework
was developped for the treatement of phonon mediated
superconductivity, its applicability does not depend on the nature
of the exchanged bosons, provided the energies of those bosons
are low compared to the electronic energies and the applicability
of Migdal's theorem is preserved. That is why in the following
we will discuss about phonons or bosons indiferently.

The gap function is proportional to the anomalous Green's function
and therefore it is intimately related with the order
parameter and the nature of the superconducting state.
Within the Eliashberg theory framework this function is the
solution of the Eliashberg equations \cite{SParks,Crev}.
For the solution of these
equations we follow the
Marsiglio Shossmann and Carbotte method \cite{MSC}. We
first solve the Eliashberg equations for imaginary Matsubara energies
$$
\Delta(i\omega_n)Z(i\omega_n)=\pi T\sum_m \Biggl[
\int_0^{\infty}d\Omega{2\Omega\alpha^2(\Omega)F(\Omega) \over
\Omega^2+(\omega_n
-\omega_m)^2}
-\mu_c\Theta(\omega_c-|\omega_m|)\Biggr]
{\Delta(i\omega_m) \over \sqrt{\omega_m^2+\Delta^2(i\omega_m)}}
\eqno(1a)
$$
$$
Z(i\omega_n)=1+{\pi T \over \omega_m}\sum_m
\int_0^{\infty}d\Omega{2\Omega\alpha^2(\Omega)F(\Omega) \over
\Omega^2+(\omega_n
-\omega_m)^2}
{\omega_m \over \sqrt{\omega_m^2+\Delta^2(i\omega_m)}}
\quad\quad\qquad \eqno(1b)
$$
and then we continue analytically to real energies by solving iteratively
the following system of equations
$$ $$
$$
\Delta(\omega,T)Z(\omega,T)=\pi T\sum_m
\Biggl[
\int_0^{\infty}d\Omega{2\Omega\alpha^2(\Omega)F(\Omega)\over
\Omega^2-(\omega-i\omega_m)^2}
-\quad\quad\qquad
\qquad\qquad\qquad
\quad\quad\qquad
$$
$$ $$
$$
\quad\quad\qquad
\quad\quad\qquad
\qquad\qquad\qquad
-
\mu^*(\omega_c)\Theta(\omega_c-|\omega_m|)\Biggr]
{\Delta(i\omega_m,T)\over
\sqrt{\omega_m^2+\Delta^2(i\omega_m,T)}}+
$$
$$ $$
$$
+
i\pi\int_0^{\infty}d\Omega\alpha^2(\Omega)F(\Omega)
\Biggl\{\Bigl[
N(\Omega,T)+f(\Omega-\omega,T)\Bigr]{\Delta(\omega-\Omega,T)\over
\sqrt{(\omega-\Omega)^2-\Delta^2(\omega-\Omega,T)}}+
\quad\quad\qquad
$$
$$ $$
$$
\quad\quad\qquad
\quad\quad\qquad
\quad\quad\qquad
+
\Bigl[N(\Omega,T)+f(\Omega+\omega,T)\Bigr]
{\Delta(\omega+\Omega,T)\over
\sqrt{(\omega+\Omega)^2-\Delta^2(\omega+\Omega,T)}}\Biggr\}
\eqno(2a)
$$
$$ $$
$$
Z(\omega,T)= 1+{i\pi T\over\omega}\sum_m
\int_0^{\infty}d\Omega{2\Omega\alpha^2(\Omega)F(\Omega)\over
\Omega^2-(\omega-i\omega_m)^2}
{\omega_m \over
\sqrt{\omega_m^2+\Delta^2(i\omega_m,T)}}+
\qquad\qquad\qquad
$$
$$ $$
$$
+
{i\pi\over\omega}
\int_0^{\infty}d\Omega\alpha^2(\Omega)F(\Omega)
\Biggl\{\Bigl[N(\Omega,T)+f(\Omega-\omega,T)\Bigr]
{\omega-\Omega \over
\sqrt{(\omega-\Omega)^2-\Delta^2(\omega-\Omega,T)}}+
\qquad
$$
$$ $$
$$
\quad\quad\qquad
\qquad\qquad\qquad
+
\Bigl[N(\Omega,T)+f(\Omega+\omega,T)\Bigr]
{\omega+\Omega\over
\sqrt{(\omega+\Omega)^2-\Delta^2(\omega+\Omega,T)}}\Biggr\}
\eqno(2b)
$$
$$ $$
where
the functions $N(\Omega,T)$ and $f(\Omega,T)$ are the statistical functions
of bosons and fermions respectively and $\Delta(i\omega_m,T)$ is the solution
of equations (1). The electron-boson coupling spectral function, or
Eliashberg function $\alpha^2(\Omega)F(\Omega)$, is related to the
coupling strength $\lambda$ by
$$
\lambda=\int_0^{\infty}{2\alpha^2(\Omega)F(\Omega)d\Omega\over\Omega}.
\eqno(3)
$$

The solution of equations (1) and (2) provides the complex gap function
$$
\Delta(\omega,T)=\Delta_1(\omega,T)+i\Delta_2(\omega,T)
\eqno(4)
$$
with excellent precision at any temperature regime \cite{Crev}.
This allows a detailled analysis of the structure of the
gap functions.
The superconducting gap $\Delta_g$ is related to
the real part of the gap function
by
$$
\Delta_g(T)=
Re\Bigl\{\Delta(\omega=\Delta_g(T),T)\Bigr\}
\eqno(5)
$$

In order to isolate the coupling effects
and avoid interference with spectral effects we
begin with the analysis of a single Einstein frequency for the boson
spectrum defined by
$$
\alpha^2(\Omega)F(\Omega)={\lambda
\Omega_E\over 2}\delta(\Omega-\Omega_E)
\eqno(6)
$$
We then consider the solutions of Eliashberg equations
for different couplings and temperature regimes,
with emphasis on the region of coupling where
qualitative deviations from the conventional BCS behavior begin to appear.

\vskip 0.4cm
\begin{center}
{\bf II.a The zero temperature behavior}
\end{center}
\vskip 0.2cm

Some of our resulting complex gap functions
at the zero temperature regime are shown in figure (1).
We can clearly identify the sharp structures associated with the
Einstein boson peak. The real part of the gap function $\Delta_1(\omega,T)$
is a measure of the effective boson mediated
quasiparticle-quasiparticle interaction.
The behavior of this function for
$\lambda=1$ (figure 1a) is essentially conventional. For frequencies lower than
$\Omega_E+\Delta_g$ it is positive indicating that the effective electron
interaction is attractive, while for higher energies it is negative
indicating a repulsive  effective interaction. If one considers that the
mediating bosons are phonons, as it is the case in conventional
low-$T_c$ superconductors, this behavior is associated to the fact that the
lattice  can only be polarized in phase by charge fluctuations with
energies lower or of the order of its characteristic energies.
The pronounced peak at $\omega=\Omega_E+\Delta_g$ is due to the resonant
exchange of phonons at this frequency. It is interesting to note
that multiphonon processes
are also visible.

For larger couplings
(figures 1b and 1c)
the remarkable point
on the $\Delta_1(\omega,T=0)$ behavior is that the multiphonon
processes become dominant for superconductivity.
In fact,
the largest
value of $\Delta_1$ corresponds to such processes.
Associated to this is the fact that $\Delta_1(\omega,T)$ remains
positive at energies which are several times
higher than the characteristic
boson energies $\Omega_E$. In the case of phonon
mediated superconductivity this should indicate that for
sufficiently strong couplings the lattice is polarized in phase
even if the energies of the charge fluctuations are
much higher than its characteristic energies.

Another important point is that
for energies lower than $\Omega_E+\Delta_g$
the gap function is essentially structureless.
This is a fundamental characteristic of Eliashberg equations
and it is closely
related to the s-wave symmetry of the considered pairing.
In fact, considering the imaginary part of the gap function
$\Delta_2(\omega,T)$
one can see that this function is zero in this region of frequencies.
Since $\Delta_2$ is associated with the damping effects,
this means that there are no relaxation processes
in this region of frequencies or, in other terms, the
gap is sharply defined and there are no
available states inside the gap.
It is not difficult to see from Eliashberg equations that
when $\Delta_2=0$ the real part $\Delta_1$ is structureless.

When the coupling grows, $\Delta_g$ becomes comparable to the
phonon (or other boson) energies $\Omega_E$. Since
the real part of the gap function $\Delta_1(\omega,T\approx 0)$
is structureless for frequencies lower than $\Omega_E+\Delta_g$,
it remains positive at least in this energy range,
even if
the characteristic phonon frequencies are much lower.
A new scale for the variations of $\Delta_1(\omega,T\approx 0)$,
independent from the phonon frequency scale $\Omega_E$,
is introduced naturally in this way. The
dominance of multiboson processes is
{\it closely related with the considered s-wave symmetry}
of the pairing.

The importance of multiphonon processes is
also clear in the
imaginary part $\Delta_2(\omega,T\approx 0)$ of the gap function.
One can see that the maximal values of $\Delta_2(\omega,T\approx 0)$
are obtained for energies equal to the gap plus several
times the Einstein energy $\Omega_E$, indicating that the most important
virtual phonon (or other boson) emission takes place at those energies.
The dominance of multiboson processes induces a broadenning
of $\Delta_2(\omega,T\approx 0)$. The broadenning of
$\Delta_2(\omega,T\approx 0)$ is associated with the strong
{\it breakdown of Landau Fermi liquid picture} for the
virtually excited states occupied by the coupled
electrons.

When the effective interaction responsible for
superconductivity orginates from the exchange of virtual bosons,
this interaction is more important if the energy is almost conserved
during the emission or the absorbtion of the boson.
As a result the most important
configurations are those with excitations (from the Fermi level)
of the order of $\Omega_E$. The validity of the quasiparticle
picture for those virtually excited states is not necessary
within the conventional Eliashberg framework. Provided
Migdal's theorem is valid, the breakdown of the quasiparticle
picture for the virtually excited states of the coupled electrons
does not imply a non Fermi liquid behaviour in the normal state.
In fact
when electron-boson vertex corrections
are neglected within Migdal's theorem,
the self-energy effects in the off-diagonal sector of the theory
are not at all related with the self-energy effects in the
diagonal sector of the theory. The question of
the validity of Migdal's theorem and of the occurence
or not of a cross-over from BCS superconductivity
to Bose condensation, is equivalent here with the
question of whether one can
admit so important self-energy effects in the off-diagonal sector
of the theory, without consider any influence on the diagonal sector.

It is not clear whether the dominance of
multiboson processes characterising those
important self-energy effects in the off-diagonal sector,
is conceptually acceptable or if it is
simply due to an arbitrary extension
of Eliashberg theory beyond the limits of its validity.
In fact the dominance of multiphonon processes implies that
the lattice can be polarized in phase by charge fluctuations
with much higher frequencies than its characteristic
frequencies and this result seems  rather unphysical.
For the moment we just remark this point as one of the anomalies
with respect to the conventional weak coupling behavior.
We remark also that
this anomaly manifest when the gap $\Delta_g$ becomes
comparable
to the boson energies $\Omega_E$
at sufficiently strong couplings ($\lambda > 2$).

\vskip 0.4cm
{\bf II.b The finite temperature behavior}
\vskip 0.2cm

We have seen
in the previous subsection that for couplings of the order of
$\lambda\approx 3$, some important qualitative deviations from the
conventional low-temperature behavior begin to appear.
The nature of those deviations
casts doubts on the validity of
ET at those coupling regimes. We investigate here the behavior at
finite temperatures, and study in particular if
analogous anomalies occur in the temperature
dependence of the gap function. We give in figures 2 and 3 some of our results
at two charcteristic temperature regimes and for the same couplings
as in figure 1.
At finite temperatures, the qualitative behavior of the gap function changes
dramatically when the coupling grows.
Since we use in all cases an Einstein boson spectrum those
anomalies are not associated with the spectral structure and the
following discussion is of general validity.
Those changes are related to the changes in the zero temperature regime
since they appear simultaneously as the coupling grows.

Let as examine more precisely these finite temperature anomalies.
At finite temperature the structure of the gap function is complex.
We examine first the weak coupling calculations.
(figures 2a and 3a). In addition to the structures at
$\Delta_g+n\Omega_E$ present also at low temperatures,
new structures appear at energies $n\Omega_E-\Delta_g(T)$.
Those structures were identified a longtime ago \cite{SParks} as
due to the recombination processes.
In those processes the injected electron combines with a thermally
excited quasiparticle and forms a bound pair by emitting $n$ bosons.
The signature of those processes appears at energies of the order
of $n\Omega_E-\Delta_g(T)$ since adding the energy of the injected electron
$\Delta_g$ and subtracting the energy of the $n$ emitted bosons
$n\Omega_E$ gives zero, namely the energy of the bound pair.
Those processes are not present in the zero temperature
regime because thermally excited quasiparticles above the gap
are necessary. The absence of recombination
processes at low temperatures is therefore
related with the s-wave symetry of the considered pairing.
As we will see in the following
those processes play a very important role at higher coupllings,
and are in fact responsible for the anomalous
(with respect to the weak coupling BCS-like regime of ET) temperature
dependence of the gap function.

The first anomaly at higher couplings with respect to the $\lambda=1$ regime
appears in the low energy behavior
of $\Delta_1(\omega, T)$.
The low energy
behaviour is very important since it is directly related
to the superconducting gap by equation (5).
We remark that when $\lambda=1$ (figs. 2a, 3a) the real part
of the gap function is almost constant at low energies.
Then the superconducting gap $\Delta_g$ obtained from
equation (5) is almost identical to the zero energy solution of
Eliashberg equations which is of course the same in real and
imaginary energies. We can consider therefore the gap as an
equilibrium quantity since
Eliashberg equations for imaginary energies (1) are sufficient to define it.
The fact that $\Delta_1(\omega, T)$ is constant at low energies is
associated with the fact that $\Delta_2(\omega, T)$ is zero at those
energies, and therefore there are no available states
inside the gap $\Delta_g(T)$ even at finite temperatures.

At stronger couplings (figs. 2b, 2c, 3b, 3c) $\Delta_1(\omega, T)$ is
no more constant at low energies. In fact we see an important
enhancement of $\Delta_1(\omega, T)$ at low energies. This behaviour
is accompanied by a non zero value of $\Delta_2(\omega, T)$ in this
region of energies. The superconducting gap
$\Delta_g$ is no more completely
determined by the imaginary axis Eliashberg equations and it becomes
a {\it dynamic} quantity.

The non-zero value of $\Delta_2(\omega,T)$ and the rapid variation
of $\Delta_1(\omega,T)$ at low energies are principally
due to the one boson recombination
processes. In fact one can follow the evolution
with the coupling of the
peak at $\Omega_E-\Delta_g(T)$ associated with the one boson
recombination processes. When the coupling grows
the value of $\Delta_g(T)$ is more important
and therefore the recombination processes appear at lower
energies in the $\Omega_E$ units.
When the coupling is sufficiently large and the gap $\Delta_g(T)$
becomes of the order of $\Omega_E/2$ or greater the
{\it recombination processes influence the relevent}
$\omega
\leq \Delta_g(T)$ {\it region} of the gap function.
The rapid enhancement of $\Delta_1(\omega,T)$ at low energies
and the negative value of $\Delta_2(\omega,T)$
can be associated with the peaks of the one praticle recombination
processes at $\Omega_E-\Delta_g(T)$. Of course when the peak
due to the recombination processes appears in the
$\omega\leq\Delta_g(T)$ region of
$\Delta_1(\omega,T)$, the gap $\Delta_g(T)$ given from equation
(5) becomes larger
and therefore the influence of recombination processes becomes even larger
and so on.

In addition the recombination processes can also be considered
as partially responsible
for the general smoothing of the gap function for large couplings.
Looking for example in figure (2b) one can see that
the gap function is surprisingly smooth given the fact that we use an
Enstein spectrum. Thermal smearing is not sufficient to explain
this fact since for $\lambda=2$ and $T=0.7T_c$ (fig.2b) the temperature is
always low compared to the boson energies ($T\approx 0.15\Omega_E$).
In fact if $\Delta_g(T)\geq \Omega_E/2$ we have a superposition of
normal processes which appear at $n\Omega_E+\Delta_g(T)$ and
recombination processes which appear at $(n+1)\Omega_E-\Delta_g(T)$.
This superposition
together with the thermal smearing explains
the general smoothink of the gap function.

On the other hand, the imaginary axis solutions of the
Eliashberg equations remain close to the BCS
behavior
even for strong couplings.
We can see this in
figure (4) where we plot the temperature dependence
of $[\Delta(i\omega_n=0,T)]^2$
solution of equations (1) for $\lambda=3$.
For weak couplings we expect this function to vary linearly with temperature
near $T_c$. This is due to the fact that when $\lambda\rightarrow 0$
we have $\Delta_g (T)\approx \Delta(i\omega_n=0,T)$ and
in the BCS regime as well as in the moderate coupling regime of ET,
the order parameter $\Delta_g(T)$ has a critical exponent $1/2$.
This exponent is associated with the mean fild nature of
those situations.
We remark in figure 4 that this exponent is valid even for $\lambda=3$,
since $[\Delta(i\omega_n=0,T)]^2$ has indeed a linear temperature dependence
near $T_c$.

The significance of the observation of the $1/2$ exponent near $T_c$ is clear.
It is associated with the
fact that in the BCS regime (as well as in the conventional moderate
coupling
regime of ET) the range of the effective interaction or the diameter of
the pairs or the coherence length $\xi$ are infinite or very large.
As a consequence there is not any visible critical regime and
the mean field treatement, which implies the $1/2$ exponent for the order
parameter, is valid even near $T_c$.
When the gap $\Delta_g$ remains an equilibrium property
for weak couplings, this standard behavior is recovered.
But, as the coupling grows, the recombination
processes influence $\Delta_g$ which becomes now a dynamical property.
However, it has been shown that
ET describes
with excellent precision the $T\rightarrow T_c$ behavior of
the high-$T_c$ cuprates \cite{GVcrit} and that in those materials we
are precisely in a regime where the gap $\Delta_g$ associated
with the energy at which the experimental density of states is
maximal $\Delta_{exp}$ \cite{GVcrit} is not an equilibrium property but a
dynamic one.

The fact that the recombination processes induce this deviation
from the mean field behavior
is significant. This clearly indicates that the system enters into
a fluctuation critical regime at strong couplings with continuous
thermal pair breaking and pair recombination. However,
by constitution, Eliashberg
theory {\it cannot} treat correctly this fluctuation critical regime
given its mean field nature.
We expect for example the existence of preformed pairs above
$T_c$, which is impossible within ET.
It is already remarkable that this theory gives such a precise
indication for the onset of this fluctuation regime.
When the gap becomes a dynamic quantity
it is necessary to go beyond Migdal's theorem in order
to treat correctly the resulting critical fluctuation regime.

In the $T\rightarrow 0$ regime we remarked that the multiboson processes
become dominat for superconductivity when, at sufficiently
strong couplings, the gap $\Delta_g(T\approx 0)$
becomes comparable to the boson
characteristic energies $\Omega_E$. This reflects extremely
important self-energy effects in the off-diagonal sector
inducing a strong breakdown of the quasiparticle picture.
The question is whether one can admit so important
self-energy effects in the off-diagonal sector whithout any
influence on the diagonal sector. This question is equivalent to the
question of the validity of Migdal's theorem.
Here,
in the finite temperature regime, we remarked that when
$\Delta_g(T)\geq \Omega_E/2$ the recombination processes
influence the gap $\Delta_g(T)$ which becomes a dynamic quantity
indicating the development of a critical fluctuation regime.
For the correct treatement of this regime it is necessary in principle
to go beyond Migdal's theorem. In both
temperature regimes the need to go
beyond Migdal's theorem reflects the cross-over from the
BCS pairing to the Bose condensation regime. We conclude
therefore that {\it the cross-over from BCS superconductivity
to Bose condensation occurs when the gap $\Delta_g$ becomes
comparable to the characteristic boson energies.}
In other terms, Eliashberg theory is clearly sufficient
only for couplings lower or of the order of $\lambda=3$.

\vskip 0.6cm
\begin{center}
{\bf III. PHYSICAL MEANING OF THE CROSS-OVER CONDITION
$\Delta_g \approx \Omega_{ph}$}
\end{center}
\vskip 0.4cm

{}From the analysis of the behavior of the gap function at
$T\rightarrow 0$ and $T\rightarrow T_c$ we conclude that
the cross-over from BCS to
Bose condensation starts when the gap $\Delta_g$ becomes
of the order of the phonon energies $\Omega_{ph}$. Here we will arrive to
the same conclusion by a simple physical argument. In this way we will
give a simple physical interpretation to the condition
for the cross-over $\Delta_g
\approx \Omega_{ph}$.

The basic difference between the BCS and the Bose
condensation regimes relies on the way the
electrons absorb or emit the virtual phonons which carry the
attractive interaction. In the case of BCS condensation the electrons
travel during the absorbtion or emission of the virtual phonons while
in the case of Bose condensation the electrons are
immobile. When we start from the BCS limit and we
enhance the coupling through Eliashberg theory we suppose that
electrons are mobile during absorbtion or emission of the
virtual phonons.

The characteristic energy of the exchanged phonons is $\Omega_{ph}$.
The characteristic time for the absorbtion or emission of
the virtual phonon $\tau$ is related to the phonon energies
$\tau\approx \Omega_{ph}^{-1}$. During this time the paired electron
travel with the fermi velocity $v_F$ and covers
a distance $L=v_F\tau$. This distance might be related to the
coherence length $\xi$ which
in the
case of a BCS-like regime characterizes
the range of the
attractive interaction.
The gap $\Delta_g$ is directly related to the
coherence length $\xi$ by $\xi=v_F/(\pi\Delta_g)$.
Physically,
\underline{the condition $\Delta_g\leq\Omega_{ph}$
is equivalent with the condition $L\leq\xi$}.
If the distance the electron covers during the
absorbtion or emission of the virtual phonon is larger than
$\xi$ the BCS concept of the coherence length has no more
physical meaning, and this puts the limits of validity for Eliashberg theory.
It is clear that retardation effects are fundamental
for the corerct description of the intermediate regime,
in agreement with what has been argued by Zheng, Avignon and
Bennemann \cite{Avignon}.

Recently
Pistolesi and Strinati \cite{Strinati},
studying the model fermionic hamiltonian introduced by
Nozi\`eres and Schmitt-Rink \cite{Noz},
arrived to the conclusion that the natural variable
for the cross-over from BCS superconductivity to Bose
condensation is the product $k_F\xi$ and that
Cooper-pair-based superconductivity is stable against bosonization
down to $k_F\xi\approx 2\pi$. Our results are in remarkable
agreements with their conclusions. In particular their
condition for the beginning of the
cross-over from BCS superconductivity to Bose condensation
$k_F\xi\approx 2\pi$
is equivalent with our condition $\Delta_g\approx \Omega_{ph}$
as can bee easily seen from the previous discussion.

The interpretation given in Ref. \cite{Strinati} of the
Uemura plot \cite{Uemura} can be transposed in terms
of the relationship between
$\Delta_g$ and $\Omega_{ph}$ directly associated with the coupling
strength $\lambda$ within the Eliashberg framework.
Within our analysis, the materials which are close to
the dashed line $T=T_B$
in the Uemura plot (see Fig. 5)
are the materials in which the gap $\Delta_g$ is of the order of the
relevant phonon frequencies for superconductivity $\Omega_{ph}$.
Within the conventional Eliashberg framework this means that these
materials {\it have a coupling strength
$\lambda\approx 3$}. The closer the materials are to the
$T=T_B$ line the higher is the coupling. It is well known in particular that
the coupling in $Nb$ is higher than the coupling in
$Al$, and that the coupling in the
Chevrel phases is even higher \cite{Crev}.

The conclusion of Ref. \cite{Strinati}
that the line $T=T_B$ in the
Uemura plot (dashed line in Fig. 5) is
associated with the
condition $k_F\xi=2\pi$ is equivalent with the coindition
$\Delta_g\approx\Omega_{ph}$ within our analysis and
can be verified by independent experiments in the superconducting
state. In particular exploiting the gap ratio measurements
one can obtain information on the relative importance of the gap
$\Delta_g$ and the relevant phonon frequencies $\Omega_{ph}$.
A systematic study of the gap ratio spectral depedence
leads to the conclusion that in the cuprates as well as in the
fullerides the {\it gap is precisely of the
order of the relevant phonon energies}
\cite{VarPRB} and that within Eliashberg theory
in both materials the couplings are strong
$\lambda\approx 3$ \cite{VarC2}. Cuprates and fullerides
are precisely close to the $T=T_B$ line in figure 5.
The interpretation of Ref. \cite{Strinati}
to the Uemura plot and its equivalent translation
given here are therefore {\it confirmed} by the independent analysis
of Ref. \cite{VarPRB}.

In the following we will see that
some characteristic structures in the tunneling and photoemission
spectra of the oxides \underline{and} the fullerides (dip-like
and second peak structures above the gap), will give
further support to the previous analysis.
In fact, a dip structure structure above the gap
has been observed
in the density of states of cuprates and
fullerides. As we will see in the next section, this structure arises
naturally when the gap is of the order of the relevant
for superconductivity phonon
energies, and its observation in cuprates and fullerides indicates
that these materials are at the beginning of cross-over from
BCS to Bose, although of course the fermionic nature clearly prevails.
A final justification
of the interpretation of Ref. \cite{Strinati}
to the Uemura plot
(and its traduction given here), will be the eventual experimental observation
of the dip structure (and the other anomalies
described in the next section) on all the materials
which are close to the $T=T_B$ line in figure Fig. 5.
In other terms {\it we predict the existence of a dip structure
above the gap in the density of states of organic, heavy fermion and
chevrel superconductors}.

\vskip 0.6cm
\begin{center}
{\bf IV. DENSITY OF STATES AND HIGH-TEMPERATURE SUPERCONDUCTORS}
\end{center}
\vskip 0.4cm

In the previous sections we localized the cross-over from BCS
superconductivity to Bose condensation. We pointed out that
retardation effects are fundamental for the occurence of the
cross-over and we gave a physical interpretation
to the condition we derived $\Delta_g\geq \Omega_{ph}$ wnd which is in fact
equivalent to that derived in Ref. \cite{Strinati}.
{}From
the interpretation of Ref. \cite{Strinati} to the Uemura plot
\cite{Uemura} we obtained some evidence that the high-$T_c$
materials are close to this cross-over.
In this section we will try to situate
more clearly the high-$T_c$ materials with respect
to this cross-over,
and statuate on the necessity or not to go beyond
Migdal's theorem in order to understand
completely the high-$T_c$ phenomenology.
In that way we will test independently the interpretation of the
Uemura plot given in Ref. \cite{Strinati} taking advantage from the
equivalent image of this interpretation we proposed in section 3.

The anomalies discussed in
sections II.a and II.b can be visible in
the one particle excitations spectrum. The density of states of the
one particle excitations is of fundamental importance since
it is an experimentally accessible quantity.
Tunneling and photoemission experiments may give direct information on
the quasiparticle spectrum of a superconductor,
and thus they are considered
to be of great importance for the theoretical understanding of
High-$T_c$ superconductivity \cite{And1}.

Those experiments are strongly dependent on eventual
surface inhomogeneities. In particular for
$YBa_2Cu_3O_7$ there are great incertainties on the surface
stoechiometry and the bulk superconducting properties
are not easilly accessible by a surface technique.
Those problems
are amplified by the fact that the coherence length $\xi$ is very
short in these materials.
Those experimental difficulties generated great controversies
at the beginning of high-$T_c$ superconductivity in particular
concerning the existence or not of a Fermi surface or of a gap.
Now the experiments are improved and the existence
of a Fermi surface is clearly established by photoemission.
On the other hand,
the surface problems which are very important for the
study of the superconducting behavior are very greatly reduced if
one replace $YBa_2Cu_3O_7$ by $Bi_2Sr_2CaCu_2O_8$.

The more precise angle resolved photoemission (ARPES) experiments on
$Bi_2Sr_2CaCu_2O_8$ \cite{Dess,Hwu}
show clearly, with the onset of superconductivity,
a transfer of spectral weight from the gap region to
a peak at approximately $45meV$ below the Fermi energy,
which indicates the opening of the superconducting gap $\Delta_g$.
However at higher energies they show some structures which appear
to be anomalous.
They show in particular a {\it dip-like} structure which appears
for energies of the order of $80-90$ $meV$ below $E_F$,
(in all directions in \cite{Hwu} and mostly in
the $\Gamma-M$ direction in \cite{Dess}).
Data from reference \cite{Hwu} also
indicate the onset of a {\it broad band} or {\it second peak}
at higher energies,
just after the dip-like structure.

On the other hand,
completely analogous features have been reported from
tunneling experiments
\cite{Mandrus,Liu}. Furthermore, by comparing ARPES data in the $\Gamma-M$
direction from \cite{Dess} and tunneling data
from \cite{Mandrus}, one can see that
these features have not only almost the same structure,
but they also have almost the same temperature dependence.
In particular their energetic position appears to be
almost temperature independent.
This gives support to the idea that these are not
experimental artifacts, but might represent specific structures of the
quasiparticle density of states.

Up to now these structures have been considered
to represent the non-conventional
nature of $Bi_2Sr_2CaCu_2O_8$ superconductivity, and have been analyzed
by Littlewood and Varma
in the context of the Marginal Fermi Liquid
\cite{Varma}, by Anderson \cite{And2}
in the context of his model of interlayer tunneling superconductivity,
by Alexandrov and Ranninger \cite{AlRann}
in the context of a polaronic system,
and by D.Coffey and L.Coffey \cite{Coffey} as evidence of d-wave pairing.
As we will see in the following, all these exotic features
can also be reproduced and understood
in the context of
s-wave, conventional, strong
coupling,
Eliashberg theory of superconductivity \cite{VQspec,VQspec1}.

Within our framework, the density of states of quasiparticles in the
superconducting state is related to that in the normal state by
$$
{N_s(\omega,T)\over N_n(\omega,T)}=Re\Biggl\{
{\omega\over \sqrt{\omega^2-\Delta^2(\omega,T)}}\Biggr\}
\eqno(7)
$$
where $\Delta(\omega,T)=\Delta_1(\omega,T)+i\Delta_2(\omega,T)$
is the gap function, solution of the
Eliashberg equations.

Since we will discuss now qualitatively the experimentally reported
anomalous
strucures we will use a more realistic spectral function
for our calculations. We will
consider the
Eliashberg function of $Pb$,
provided by inversion of tunneling data
\cite{Rowell} (shown in figure (6))
multiplied by a constant factor $a$.
The coupling strength $\lambda$ defined by
$
\lambda=a\int_0^{\infty}2d\Omega\alpha^2(\Omega)F(\Omega)/ \Omega
$,
adjusting the parameter $a$, will enable us to consider three coupling regimes,
$\lambda=1$, $\lambda=3$ and $\lambda=5$. As will be shown later, the choice
of the spectral function will not have much effect since the
interesting structures are not
associated with spectral structures, and our qualitative discussion
is generic.
The coulomb pseudopotential will be taken equal to zero,
bearing in mind that a nonzero value of $\mu^*$ reproduces the same
phenomenology but for slightly higher values of the coupling, which depend
on the exact value of $\mu^*$.

The calculated density of states for different temperatures, and for the
cited values of $\lambda$, is reported in figure (7). For $\lambda=1$,
(Fig. 7a), the
density of states has the standard BCS behaviour.
The peak indicating the gap $\Delta_g(T)$ is clear, and the density of states
is zero for $\omega< \Delta_g(T)$. The temperature dependence of the
peak position, indicating the temperature dependence of the gap,
follows quite well the BCS law
$
\Delta(T)\approx 3.06T_c\sqrt{1-T/T_c}
$.

As we already noticed
the experimental behaviour of $Bi_2Sr_2CaCu_2O_8$
is different from this behaviour on four counts.
First, the density of states inside the gap is nonzero for
$T\neq 0$, and furthermore it rises with rising $T$. Second, the peak position
does not seem to move to zero as $T\rightarrow T_c$, third, after the peak
there is a dip-like structure already mentioned, and fourth, there is a
second peak structure.

Looking now to the stronger coupling calculations, $\lambda=3$ and $\lambda=5$,
(Figures 7b and 7c),
one can see that all four above-mentioned anomalies with respect to the
BCS phenomenology are present.
The finite density of states inside the gap at $T\neq 0$ and the anomalous
temperature dependence of the peak position for strong couplings were first
reported numerically by Allen and Rainer \cite{AlRainer}
together with the absence of the Hebel-Slichter peak in the
NMR relaxation at the same couplings.
We report here that this behavior is accompanied, at higher
and temperature independent
frequencies, by a {\it dip-like} and a {\it second-peak} structure.
The form and temperature dependence of
these structures are {\it very similar} to
the behavior of the experimentally reported anomalous structures
\cite{Dess,Hwu,Mandrus,Liu}.
We insist to the fact that the temperature independence of the
energetic position of those structures reported experimentally in Refs.
\cite{Mandrus,Dess}
is perfectly reproduced in figure (7).
One can
conclude already that {\it the quasiparticle spectrum provided
by tunneling and photoemission on $Bi_2Sr_2CaCu_2O_8$, is consistent
with strong coupling s-wave boson exchange superconductivity}.

Before going further in analyzing the physical meaning of those structures,
we can make in figure (7) another important remark.
The exact frequency position (in $\Delta_g(T=0)$ units) of
the dip-like structure and of the second peak structure are coupling
strength dependent. The higher are the couplings the lower are the
energies (in $\Delta_g(T=0)$ units) at which those structures appear.
The origin of this behavior will be discussed later.
This provides us with a fundamentally new tool in extracting the coupling
strength from experiments.
To exploit this new tool it is necessary to study systematically the
influence of the spectral structure.
The systematic exploitation of
this new tool will be left for a future publication. For the moment we
just precise here that a crude comparison of experimental data
with the present calculations leads to couplings of the order of $\lambda
\approx 2.5-4.0$ although the experimental data seem not
sufficiently precise. Therefore, our analysis is quantitatively
quantitatively compatible with the experimentally
reportad structures.

Let us analyze now carefully what is the physical meaning of those
anomalies. The first point we remark is that those structures are
\underline{not} associated with
spectral structures. To illustrate this we display
in figure (8) the density of states obtained
using an Einstein
spectrum  with $\lambda=3$ at $T=0.3T_c$. We can clearly identify
the dip-like structure and the second peak structure
to be completely independent from the
spectrum which causes sharp structures in the shorter
phonon frequency $\Omega_E$ scale. In fact as we will point out in the
following those anomalous structures are associated with
the anomalies in the gap function localized in section II,
and indicate that {\it high-$T_c$ materials
are at the beginning of a cross-over from BCS superconductivity to
Bose condensation, the fermionic character being clearly dominant}.
Notice that in the same conclusion arrived Rietveld Chen and Van der Marel
\cite{vandermarel} from a completely independent experimental
analysis.

To understand the anomalous dip-like and second peak strucrures
which appear at energies higher than the gap, it is useful to
make a $\Delta(\omega,T)/\omega\ll 1$ expansion of
equation (7), as it has been done in the
pioneering work of Scalapino Schirieffer and Wilkins
\cite{SSW}.
The reduced density of states can be written as follows
$$
{N_s(\omega,T)\over N_n(\omega,T)}\approx
1+{1\over 2}\biggl[ \biggl( {\Delta_1(\omega,T)\over \omega}\biggr)^2-
\biggl( {\Delta_2(\omega,T)\over \omega }\biggr)^2\biggr]+...
\eqno(8)
$$
The dip-like structure appears when $\Delta_1(\omega,T)$ drops
down and $\Delta_2(\omega,T)$ becomes dominant.
In the conventional weak coupling regime the energetic scale
of the variations of the gap function is determined by the
boson energies. Therefore, equation (8) illustrates why the boson spectrum
of the conventional superconductors is visible in
the tunneling experiments on conventional superconductors.

However,
as we have seen in section II.a, at strong couplings,
the scale of the variations of the gap function is
no more associated with the boson energies.
This is the reason why the dip-like and second peak structures
are  {\it not associated with the spectral structure}.
As we discussed in section II.a,
this new scale of variations of the gap function reflects very important
self-energy effects in the off-diagonal sector.
{\it The dip-like structure indicates, therefore, the breakdown of the
Landau Fermi-liquid picture beacause of the strong coupling to bosons}.

The second peak structure is also not associated with spectral structures
and follows the developpement of the dip-like structure.
Looking in figure (1c), one can see that
$\Delta_1(\omega,T)$ is positive at low energies, then changes sign
at approximatively $\Delta_g+3\Omega_E$ and then it changes for a second
time sign at approximatively $\Delta_g + 6\Omega_E$. The second peak structure
is associated precisely with this second sign changement of
$\Delta_1(\omega,T)$. This is obvious if we compare figure (1c)
with figure (8). We have seen
that $\Delta_1(\omega,T)$ is positive at
low energies because the lattice is polarized in
phase by charge fluctuations with energies
lower or of the order of its characteristic energies and the effective
electron interaction reflected by $\Delta_1(\omega,T)$ is attractive
at those energies.
It is unclear why the effective interaction after being
repulsive at higher energies,
becomes once more attractive at even higher energies.
However this behavior is clearly associated to the fact that the
gap function has now a new scale of variations.

In the $\lambda\rightarrow\infty$ and $\lambda = 100$ calculations
presented in reference \cite{MCinf}, the density of states
has a multipeaked structure with narrow peaks ({\it delta}
peaks for $\lambda\rightarrow\infty$) at energies
which are multiples of the gap energy. It has been suggested
by S. Kivelson (see Ref. 18 in reference \cite{MCinf}) that those
multiple peak structures might indicate the
local nature of the pairs in this regime.
However this $\lambda\rightarrow\infty$ regime
of Eliashberg thery is quantitatively different from a
Bose condensation regime \cite{MRR}. This is natural since the
Bose condensation regime corresponds to a zero  Fermi velocity
and therefore Migdal's theorem has no more any sense.

Although Eliashberg theory is certainly not valid in the $\lambda\rightarrow
\infty$ regime considered in \cite{MCinf}, it is absolutely not clear
if it is so in the much more moderate regime $\lambda\approx 3$
considered here. We think nevertheless, that the picture proposed
in Ref. \cite{MCinf}
for the multipeaked structure might be relevant for our second
peak structure. Indeed the second peak structure might be associated
with the presence
at  $\lambda\approx 3$ of almost localized pairs.
Within this picture our results indicate {\it
the coexistence in the high-$T_c$ materials of BCS pairing and
almost localized pairs}. Apparently it is possible to have both
BCS excitations breaking the superconducting order and
``individual'' excitations of almost localized pairs.

The association of the second peak structure with the existence of
almost localized pairs is in agreement with the presence of this
peak just after the dip-like structure. We have seen in fact that
the dip-like structure indicates the strong breakdown of quasiparticle picture
for the states occupied by the paired electrons.
In the BCS limit where the states occupied by the coupled electrons are
free electron states (plane waves),
the pairs are extended to infinity ($\xi \rightarrow\infty$).
The important self energy effects reflected by the dip-like structure
result to an important limitation of the coherence length $\xi$.
In the high-$T_c$ materials in particular, the coherence length $\xi$
is of the order of the lattice spacing \cite{Bgg}, and thus the
presence of almost localized pairs is not unreasonnable.
Therefore,
dip-like structure and second peak structure are perfectly compatible
and both reflect the very important
self-energy effects in the off-diagonal sector
due to the strong electron-boson coupling.

Associating the second peak with the presence of almost localized pairs,
we are also able to understand
why this structure appears at lower energies (in $\Delta_g(T=0)$ units)
when the coupling grows. In fact the second peak structure must appear
at energies higher than $3\Delta_g$, because $\Delta_g$ is the energy
of the injected electron and for the
excitation of a completely localized pair (with $\xi\rightarrow 0$)
one needs an
additional energy of the
order of $2\Delta_g$ \cite{MRR,MCinf}.
However the pairs are not completely
localized in the coupling regime that we consider here ($\xi\neq 0$)
that is why we call them almost localized pairs. For the excitation of
an almost localized pair we need an energy higher than $2\Delta_g$
since one has to take into account the superposition
with the other almost localized pairs. When the coupling grows
the spatial extension of the pairs and also the superposition
of the pairs is reduced.
As a result,
when the coupling grows,
the second peak appears at lower energies in
$\Delta_g$ units reflecting precisely
the reduction of the superposition of the almost localized pairs.
We understand, therefore, why the energetic position (in
$\Delta_g(T=0)$ units) of the anomalous dip-like and second-peak
structures can be an interesting new tool for extracting the coupling
strength from experiment.

Concerning the temperature independence of the energetic position
of those structures,
we can understand its origin looking in
figures (1c) and (3c). Comparing those last figures one can see that
for couplings of the order of $\lambda\approx 3$ the gap function
has the same energetic scale for its variations
at $T=0.3T_c$ (figure 1c) and  $T=0.9T_c$ (figure 3c).
The temperature independence of the energetic position
of the dip-like and second peak structures, reflects the
temperature independence of the energetic scale of the variations
of the gap function.

We already remarked that
the new scale of the variations of the gap function
appears when the gap $\Delta_g$ becomes comparable to the boson
energies. Since the scale of variations of the gap function
is then influenced by the gap $\Delta_g$ its temperature independence
might be associated with the anomalous temperature behaviour
of $\Delta_g(T)$ reflected by the first peak in the density
of states. Therefore, the anomalous temperature
behaviour at low energies ($\omega\leq
\Delta_g$)
of the density of states first reported in \cite{AlRainer}
is intimately related to the
temperature independence of the energetic
position of the "anomalous"
dip-like and second-peak structures at higher energies.

We have seen
previously that the whole "anomalous"
behavior at high energies of the density
of states can be associated with the
high energy behavior
of the gap function at those strong couplings. It is natural
to expect that the
anomalous temperature dependence
of the density of states at low energies might
reflect the anomalous temperature behavior of the gap function
at lower energies. We have discussed this behaviour in
section II.b. We have seen in particular that when
$\Delta_g(T)\geq \Omega_E/2$ the recombination processes influence the
relevant $\omega\leq \Delta_g$ region in the density of states
indicating that the system developps a critical fluctuation regime.
As a result the gap is no more a thermodynamic quantity but
becomes a dynamic quantity.

The anomalous temperature dependence of the first peak in the density
of states (indicating the anomalous temperature dependence of
$\Delta_g(T)$ discussed in section II.b) and the finite density
of states inside the gap are due to the {\it recombination processes}
and they also indicate that {\it highh-$T_c$ materials are at the beginning of
a cross-over from BCS superconductivity to Bose condensation}.
We remark at that point
that using a phenomenological parametric approach,
Dynes et al. \cite{Dynes} indentified recombination processes
as responsible for the much weaker damping effects in
the density of states of
low temperature superconductors.
As we discussed in section II.b, the dominance of those processes
is evidence of a cross-over only when it concerns the low energy
region (gap region $\omega\leq \Delta_g(T)$). This occurs
at sufficiently strong couplings ($\lambda> 2$)
and in the high-$T_c$ superconducting state.

\vskip 0.6cm
\begin{center}
{\bf V. UNDERSTANDING THE ROBUSTNESS OF ELIASHBERG THEORY BY
INCLUDING
CORRECTIONS BEYOND MIGDAL'S THEOREM}
\end{center}
\vskip 0.4cm

In the previous section we have seen that high-$T_c$ materials
are at the beginning of a cross-over from BCS superconductivity
to Bose condensation. However Eliashberg theory appears surprisingly
robust in the description of the phenomenology of those materials
and therefore appears quantitatively relevant in a regime
where in principle a generalization beyond Migdal's theorem
should be in principle necessary.

Recently there have been detailled studies of the
first order non-adiabatic corrections to Migdal's theorem
\cite{LP,LP2}. In particular the Eliashberg equations
can be generalized near $T_c$ in the following
way \cite{Pat}:
$$
\Delta(i\omega_n)Z(i\omega_n)=
\pi T_c \sum_m {\lambda_\Delta (i\omega_n,i\omega_m;Q_c)
\Omega_E^2\over (\omega_n-\omega_m)^2+\Omega_E^2}
{\Delta (i\omega_m)\over |\omega_m|}
{2\over\pi}arctg\Biggl({E\over 2Z(i\omega_m)|\omega_m|}\Biggr)
\eqno(9)
$$
$$
Z(i\omega_n)=1+{\pi T_c\over \omega_n}
\sum_m {\lambda_Z (i\omega_n,i\omega_m;Q_c)
\Omega_E^2\over (\omega_n-\omega_m)^2+\Omega_E^2}
{\omega_m\over |\omega_m|}
{2\over\pi}arctg\Biggl({E\over 2Z(i\omega_m)|\omega_m|}\Biggr)
\eqno(10)
$$
where $\Omega_E$ is the energy of the considered phonon
Einstein spectrum, and the adiabatic corrections were included
in the effective coupling functions $\lambda_\Delta$ and $\lambda_Z$
defined as follows
$$
\lambda_\Delta (i\omega_n,i\omega_m;Q_c)= \lambda\bigl[
1+2\lambda P_V (i\omega_n,i\omega_m;Q_c)+\lambda P_c
(i\omega_n,i\omega_m;Q_c)\bigr]
\eqno(11)
$$
$$
\lambda_Z(i\omega_n,i\omega_m;Q_c)=
\lambda\bigl[ 1+\lambda P_V (i\omega_n,i\omega_m;Q_c)\bigr]
\eqno(12)
$$
The functions $P_V$ and $P_c$ refer to the first order
non-adiabatic corrections in the diagonal and off-diagonal sectors
respectively and $\lambda$ is the electron-phonon
coupling in the conventional Eliashberg theory framework.
It is supposed a momentum independent scattering up to a momentum
cut-off $Q_c=q_c/2k_F$ and in the case of half-filling $E=2E_F$.
Details on the considered model and the approximations made
to obtain the first non-adiabatic corrections
$P_V$ and $P_c$ and on their
detailled structure
can be found in Ref. \cite{LP2}.

A very important remark has been done on the
behavior of the physical solutions of euqations (9-12).
It has been pointed out in Ref. \cite{Pat}
that the frequency structure of $P_V$ and $P_c$ is marginal
compared in particular with the momentum cut-off dependence
of $T_c$ obtained solving
equations (9-12). Neglecting the dependence of the non-adiabatic
corrections $P_V$ and $P_c$ on the Matsubara frequencies and
replacing them by their average over those frequencies,
leads with a good precision to the same $T_c$
as that obtained if one take into account the full frequency
dependence \cite{Pat}.

When
the $\omega_m$ dependence is neglected, equations (9-12)
{\it have a form very similar to that of the conventional
Eliashberg equations} (Eqs. 1a and 1b) for $T\rightarrow T_c$.
This is more clear for weak couplings where the renormalization function
$Z(i\omega_n)$ can be replaced by the unity in the module
of arctg in equations (9) and (10). Then the non-adiabatic
corrections play the role of {\it effective couplings}, which nevertheless
have a strong dependence on the momentum cut-off $Q_c$ and on the
Migdal parameter $m=\Omega_E/E_F$, and maybe different for different
properties \cite{Pat}.
Unfortunately it
is rather difficult to write generalized Eliashberg equations at
$T\rightarrow 0$. This should be very interesting
since one could maybe give a more complete interpretation
to the condition for the cross-over $\Delta_g(T=0)\approx \Omega_{ph}$,
introducing both momentum and adiabaticity effects.
Work in this direction is in progress.

The fact that the first order non-adiabatic corrections
act as an effective coupling, gives a clue to understand
{\it the robustness} of Eliashberg theory in the description
of the superconducting state even when the cross-over from
Eliashberg theory to Bose condensation has started. This
indicates that the
corrections beyond Migdal's theorem
should have more qualitative effects in the normal state than in the
superconducting state. If we examine the high-$T_c$ phenomenology
we remark that this is precisely the case, and therefore
the detailled study of such corrections might be an important
step for the understanding of some anomalies in the
normal state phenomenology.

\vskip 0.6cm
\begin{center}
{\bf VI. CONCLUSIONS}
\end{center}
\vskip 0.4cm

We studied carefully the Eliashberg theory behavior in the coupling
region where some important qualitative deviations from the
BCS behavior appear. Such deviations are:

-Dominance of multiphonon processes for superconductivity;

-Dissociation of the scale of variations of the gap function from the
phonon energetic scale;

-The superconducting gap $\Delta_g$ is no longer
a thermodynamic quantity but becomes a dynamic one;

-The energetic scale of the variations of the gap function
is temperature independent;

We showed that
all these qualitative deviations occur when $\Delta_g$ becomes of the
same order with the characteristic energies of the phonons
(or other bosons in ``exotic'' theories) which mediate the
attractive interaction $\Omega_{ph}$, and indicate the beginning
of a cross-over from BCS superconductivity to Bose condensation.
We pointed out that the condition $\Delta_g\approx \Omega_{ph}$
is in fact equivalent to the condition $k_F\xi\approx 2\pi$
derived in Ref \cite{Strinati} and also to the physical constraint
$L\approx \xi$ where $L$ is the distance the paired electron covers
during the absorbtion of the virtual phonon and $\xi$ the
superconducting coherence length.


The previous anomalies are reflected by anomalies in the
density of states of excitations which is an experimentally accessible
quantity. Such anomalies are:

-Finite density of states inside the gap at finite temperatures;

-Enhancement of the density of states inside the gap by enhancing the
temperature;

-The peak indicating the experimental gap do not move to zero
when $T\rightarrow T_c$;

-Dip structure above the gap at temperature independent but
coupling strength dependent energies;

-Second peak structure above the dip which
has the same temperature and coupling
dependence with the dip structure;

We showed that the finite density of states inside the gap
and the whole anomalous temperature behavior of the density of states
is due to the recombination processes. The anomalous dip and second peak
structures indicate the breakdown of Fermi liquid picture
for the virtually excited states occupied by the paired electrons,
because of the strong electron-phonon coupling.

The interpretation done in Ref. \cite{Strinati} of the
Uemura plot \cite{Uemura} is confirmed from our
analysis. Taking into account the analysis of Ref. \cite{Strinati}
{we predict the presence of a dip structures
(and all the previously cited anomalies) in all
the materials that are close to the $T=T_B$ line in
figure 5 (cuprates, fullerides, organic superconductors, heavy fermion
compounds, chevrel phases etc.)}
Since within Eliashberg theory the cross-over
starts at couplings of the order of
$\lambda > 2$ (for which $\Delta_g\approx\Omega_{ph}$),
if we refer to the experimental gap ratios
$2\Delta_g(T\approx 0)/T_c$ od cuprates and fullerides
\cite{VarPRB,VarC2}, we have another independent
evidence that the high-$T_c$ materials might be
concerned from the previously cited anomalies.
Indeed,
tunneling and photoemission experiments on cuprates
\cite{Dess,Hwu,Mandrus,Liu,Kita} and fullerides \cite{Knupfer}
appear in perfect agreement with the previous predictions.
{\it We predict a dip structure in organic and heavy fermion
superconductors}. The eventual observation of the dip in these last
materials will definetely confirm our analysis.

Another important point emerging
from the comparison of our analysis with the experimental behavior of
cuprates, is the surprising robustness of
Eliashberg theory at least concerning the one-particle excitation spectrum
considered here. In fact the critical fluctuation regime \cite{GVcrit}
and the anomalously large self-energy effects in the
off-diagonal sector resulting in the coexistence of BCS pairing
and almost localized pairs, are situations for which the
generalisation
of Eliashberg theory beyond Migdal's theorem appears
inavoidable. The robustness of our framework
is understood considering a generalization of Eliashberg equations
to include first order non-adiabatic corrections \cite{LP,LP2,Pat}.
The frequency structure of the aconsidered
corrections, is for some properties irrelevant, in which
case the generalized Eliashberg equations have a similar
structure with the bare-ones, except that the effective
couplings have an internal momentum structure
and can be different
for different properties. Since the first non adiabatic corrections
do not have significant influence on the frequency structure
of Eliashberg equations, they do not introduce significant qualitative
effects on the resulting phenomenology. Therefore, the
bare equations remain qualitatively relevant at the beginning of the cross-over
regime.

On the other hand,
non-adiabatic corrections
can have important even qualitative
implications on other parts
of the phenomenology. This is more probable in the dynamical
properties and especially in the normal state
infrared conductivity.
In fact while the low energy dynamic behaviour in the
high-$T_c$ superconducting state appears to be evidence for
the relevance of Eliashberg theory \cite{GVPLA}
and the behaviour near $T_c$
quantitatively understood within ET \cite{GVcrit},
the temperature dependence of the infrared conductivity in the
normal state is difficult to accomodate within this framework
\cite{Rprive}.
In general we
expect the deviations from ET induced by those corrections more important
in the normal state.
If we consider
for example the NMR relaxation rate measurements,
although the absence of the Hebel-Slichter peak in the superconducting
state due  to a developement of a critical fluctuation regime
\cite{GVcrit}
can be understood within ET
\cite{AlRainer,GVcrit,CV,GVPLA},
for the understanding of deviations from the Korringa
behavior
in the normal state
(which should be due to the existence of
preformed pairs above $T_c$) it is necessary to consider
non-adiabatic vertex corrections.

In conclusion our results indicate that high-$T_c$ materials are
at the {\it beginning} of a cross-over from BCS superconductivity to
Bose condensation, the BCS component being dominant.
Retardation effects are fundamental for the occurance
of this cross-over and therefore might be
included in any attempt to describe
the intermediate regime \cite{Avignon}.
The Eliashberg conventional framework appears to be rather
robust eSpecially concerning the one particle behavior
as well the low energy dynamics
in the superconducting state.
The eventual deviations of the high-$T_c$ phenomenology
from the Eliashberg behavior (especially in the normal state)
might be due to the
impossibility of this theory to describe completely
this cross-over regime. The study of non adiabatic vertex corrections
might be fundamental for the understanding of those deviations.

\vskip 0.6cm
\begin{center}
{\bf Acknowledgements}
\end{center}
\vskip 0.4cm

It is a pleasure to acknowledge valuable discussions with
Prof. B. Chakraverty,
Prof. G.M.Eliashberg, Prof. D.Rainer
and Prof. G.C.Strinati.
We also acknowledge support from the Human Capital and Mobility
program of the CEE under contract ERBCHBICT930906.

\newpage

\newpage
\begin{centerline}
{\bf FIGURE CAPTIONS}
\end{centerline}
\vskip 1.2cm

{\bf Figure 1:} The real part (full lines)
and the imaginary part (dashed lines) of
the gap function solution of the Eliashberg equations
using an Einstein spectrum at the ``zero'' temperature
regime $T=0.3T_c$ and for couplings: (a)
$\lambda=1$, (b) $\lambda=2$, and (c) $\lambda=3$.
\vskip 0.9cm

{\bf Figure 2:} Same as in figure 1 for the finite
temperature regime $T=0.7T_c$. The couplings considered
are: (a) $\lambda=1$, (b) $\lambda=2$, and (c) $\lambda=3$.
\vskip 0.9cm

{\bf Figure 3:} Same as in figure 2 for $T=0.9T_c$
\vskip 0.9cm

{\bf Figure 4:} The temperature dependence of the
square of the
gap function solution of equations
(1) for a zero imaginary energy. Notice the linear temperature dependence
near $T_c$.
\vskip 0.9cm

{\bf Figure 5:} The Uemura plot (fig. 3 in Ref. \cite{Uemura}).
We predict a dip structure above the gap (and various other anomalies
described in section IV) in the density of states of
all the materials that are close to the $T=T_B$ line.
In these materials, the superconducting gap $\Delta_g$ is of the same order
with the characteristic energies of the boson mediators of
superconductivity (in conventional theories the relevant phonon energies).
\vskip 0.9cm

{\bf Figure 6:} The Eliashberg function of $Pb$ used in the
calculations of figure 7.
\vskip 0.9cm

{\bf Figure 7:} The reduced density of states $N_s(\omega,T)/N_n(\omega,T)$
for the $Pb$ spectrum (shown in figure 5),
at different coupling regimes, and for temperatures
$T=0.3T_c$ (full), $T=0.7T_c$ (dotted), $T=0.9T_c$ (dashed), $T=0.95T_c$
(dot-dashed), and $T=0.975T_c$ (triple-dot-dashed).
\vskip 0.9cm

{\bf Figure 8:} The reduced density of states obtained using an
Einstein spectrum with $T=0.3T_c$ and $\lambda=3$. The corresponding
gap function is shown in figure 1c.
\vskip 0.9cm

\end{document}